# Deep Learning Approaches for Network Traffic Classification in the Internet of Things (IoT): A Survey

Jawad Hussain Kalwar
dept. of Software Engineering
Mehran University of Engineering and Technology
Jamshoro, Sindh, Pakistan
22MESE12@students.muet.edu.pk

Sania Bhatti
dept. of Software Engineering
Mehran University of Engineering and Technology
Jamshoro, Sindh, Pakistan
sania.bhatti@faculty.muet.edu.pk

**ABSTRACT**

The Internet of Things (IoT) has witnessed unprecedented growth, resulting in a massive influx of diverse network traffic from interconnected devices. Effectively classifying this network traffic is crucial for optimizing resource allocation, enhancing security measures, and ensuring efficient network management in IoT systems. Deep learning has emerged as a powerful technique for network traffic classification due to its ability to automatically learn complex patterns and representations from raw data. This survey paper aims to provide a comprehensive overview of the existing deep learning approaches employed in network traffic classification specifically tailored for IoT environments. By systematically analyzing and categorizing the latest research contributions in this domain, we explore the strengths and limitations of various deep learning models in handling the unique challenges posed by IoT network traffic. Through this survey, we aim to offer researchers and practitioners valuable insights, identify research gaps, and provide directions for future research to further enhance the effectiveness and efficiency of deep learning-based network traffic classification in IoT.

## 1 INTRODUCTION

The Internet of Things (IoT) has the prospect to revolutionize the way we interact with the digital world, connecting billions of devices and enabling numerous applications across various domains such as healthcare, smart cities, transportation, and industry. However, the rapid growth of IoT devices has also brought about significant challenges, particularly in managing and securing the vast amount of network traffic generated by these devices[62]. Network traffic classification plays a crucial role in understanding the communication patterns, identifying potential security threats, and optimizing network performance in IoT environments[22]. Traditional approaches to traffic classification, such as port-based or payload-based methods, have limitations in effectively handling the diverse and dynamic nature of IoT traffic[1]. Consequently, researchers have turned to deep learning techniques, which have demonstrated remarkable success in various domains, including computer vision, natural language processing, and speech recognition[49].

This survey paper aims to present a comprehensive analysis of the state-of-the-art deep learning approaches employed for network traffic classification in the context of the IoT. By accumulating existing literature and identifying emerging trends, this survey contributes to the understanding and advancement of the field. Furthermore, it serves as a valuable resource for researchers, practitioners, and network administrators seeking to navigate the complex landscape of network traffic classification in IoT environments.

The primary objective of this survey paper is to provide a review and analysis of the deep learning approaches utilized for network traffic classification in the IoT. Specifically, we aim to:

- Identify and tabulate the various deep learning models and architectures employed in network traffic classification for IoT environments.

- Investigate the methodologies and techniques used for feature extraction, representation, and selection in deep learning-based traffic classification.

- Analyze the strengths, limitations, and performance of existing deep learning methods in this domain.

- Highlight the challenges and open research directions in deep learning-based network traffic classification for IoT.

The scope of this survey encompasses recent literature published in academic journals, conferences, and reputable online repositories since January 2020 up until April 2023. We focus on the application of deep learning techniques to network traffic classification in IoT environments, considering both supervised and unsupervised learning approaches. However, we exclude studies that solely focus on traditional machine learning methods or other specific non-deep learning techniques. We have also excluded the studies that are not specifically aimed at IoT and the unique challenges that come with that.

The remainder of this survey paper is organized as follows. Section 2 provides an overview and the Background concepts of the fundamentals of network traffic classification in the IoT, including the characteristics of IoT traffic and the challenges faced in its classification. Section 3 presents a brief look at related surveys. Section 4 describes our methodology. Section 5 presents a comprehensive survey of deep learning models and architectures used in IoT network traffic classification, highlighting their advantages and drawbacks. It discusses the methodologies employed for feature extraction, representation, and selection in deep learning-based traffic classification. the challenges and open research directions. followed by a summary and conclusion. By addressing these key aspects, this survey paper aims to shed light on the current landscape, achievements, and challenges in the application of deep learning for network traffic classification in the IoT domain.

## 2 BACKGROUD

Deep learning has emerged as a powerful technique for network traffic classification in the Internet of Things (IoT) domain[1]. With the exponential growth of connected devices, the need for efficient and accurate identification of network traffic becomes crucial for managing and securing IoT systems. Deep learning models, particularly convolutional neural networks (CNNs)[2,17,28,41] and recurrent neural networks (RNNs)[6,27,37,39,50], have demonstrated remarkable capabilities in extracting relevant features from raw network data and effectively classifying different types of traffic. By leveraging the inherent ability of deep learning models to learn hierarchical representations, IoT network traffic can be effectively classified based on its distinctive patterns and characteristics.

Furthermore, the availability of large-scale IoT traffic datasets[3,33,36], enables the training of deep learning models with enhanced accuracy and generalizability. However, challenges persist in developing deep learning models that are computationally efficient and resource-constrained for IoT devices[2]. Despite these challenges, deep learning remains a promising approach for network traffic classification in IoT, offering significant potential for improving network management, security, and resource allocation in IoT environments[22].

Network traffic classification, an essential task in network management and security, involves categorizing network traffic into different types or classes based on various attributes and characteristics[8]. The primary motivation behind network traffic classification is to gain insights into network behavior, monitor performance, allocate resources effectively, and ensure network security. By understanding the composition and patterns of network traffic, administrators can identify and prioritize critical applications, detect anomalies, mitigate network congestion, and enforce security policies[24]. Furthermore, network traffic classification enables the development of specialized services, such as Quality of Service (QoS) mechanisms and intrusion detection systems, tailored to different traffic types[32]. Effective network traffic classification is crucial for managing and optimizing network resources, enhancing user experience, and safeguarding network infrastructure from potential threats and vulnerabilities.

The Internet of Things (IoT) refers to the interconnected network of physical devices embedded with sensors, actuators, and communication capabilities that enable them to exchange data and interact with the environment. These devices, ranging from everyday objects like smart appliances and wearables to industrial machinery and infrastructure, collect and transmit data to the cloud or other networked systems[40]. Network traffic classification in the IoT context involves the categorization of data flows generated by IoT devices based on their specific characteristics and purposes. The unique nature of IoT network traffic presents several challenges compared to traditional network traffic classification[42]. First, the sheer scale and diversity of IoT devices result in a significantly larger volume of network traffic with distinct patterns and behaviors. This necessitates the development of classification techniques capable of handling the increased complexity and variability in IoT data[29]. Second, IoT devices often operate under resource constraints, such as limited processing power, memory, and energy. As a result, network traffic classification algorithms for IoT must be optimized to minimize computational overhead and energy consumption[44]. Moreover, IoT traffic may exhibit time-varying patterns, intermittent connectivity, and mobility, requiring adaptive classification methods that can handle dynamic and evolving network environments[31]. Additionally, IoT network traffic classification may involve the identification of specific IoT applications, protocols, or device types to enable targeted monitoring, resource allocation, and security measures. Therefore, network traffic classification in the IoT domain requires tailored algorithms and techniques that can handle the unique characteristics and challenges posed by the vast and diverse IoT ecosystem[9]. Researchers are actively exploring novel approaches, including deep learning, machine learning, and data mining techniques, to tackle these challenges and advance the state-of-the-art in IoT network traffic classification.

Machine learning refers to the field of study and practice that focuses on the development of algorithms and models capable of automatically learning patterns and making predictions or decisions based on data. It encompasses various techniques such as statistical models, decision trees, and neural networks[23]. Deep learning, a subset of machine learning, specifically refers to the use of artificial neural networks with multiple layers to extract high-level representations from raw data[46]. Deep learning excels in capturing intricate patterns and dependencies in complex datasets, enabling more accurate and robust predictions compared to traditional machine learning methods. The hierarchical nature of deep neural networks allows them to automatically learn abstract features and representations, making them highly effective for tasks like image recognition, natural language processing, and speech recognition[12]. Consequently, deep learning has gained prominence due to its superior performance on large-scale datasets, ability to handle unstructured data, and the potential for end-to-end learning without explicit feature engineering. In summary, deep learning's capacity to capture complex relationships and learn hierarchical representations makes it a powerful and superior approach compared to vanilla machine learning for many data-driven tasks.

Deep learning has emerged as a promising approach for network traffic classification in the Internet of Things (IoT) domain, offering the potential to effectively manage and secure IoT systems[22]. The application of deep learning to network traffic classification involves a multi-step process. First, a large-scale dataset of labeled network traffic is collected, representing various IoT applications, protocols, and device types. This dataset serves as the training data for the deep learning model. Next, the dataset is preprocessed to extract relevant features, such as packet headers, payload information, flow statistics, and timing characteristics. These features capture the distinctive patterns and behaviors of different network traffic classes[59]. Then, a deep learning model, typically a convolutional neural network (CNN)[2,17,28,41] or recurrent neural network (RNN)[6,27,37,39,50] is constructed. The model architecture consists of multiple layers, allowing it to learn hierarchical representations and abstract features from the raw network traffic data. The model is trained using the labeled dataset, employing techniques like backpropagation and stochastic gradient descent to optimize the model's parameters. Once trained, the deep learning model can classify incoming network traffic into predefined classes or detect anomalies based on learned patterns[7].

Compared to traditional methods of network traffic classification, deep learning offers several advantages. First, deep learning models can automatically learn and extract complex and hierarchical features from raw data, eliminating the need for manual feature engineering. This ability is particularly valuable in the IoT context, where the characteristics of network traffic can be diverse, dynamic, and constantly evolving. Second, deep learning models are highly flexible and adaptable, capable of capturing both spatial and temporal dependencies in network traffic data. This flexibility enables the classification of time-varying and intermittent IoT traffic patterns. Third, deep learning models have shown superior performance in handling large-scale datasets, which is crucial in IoT environments with a vast number of connected devices generating massive amounts of data. Furthermore, the availability of pre-trained models and transfer learning techniques allows for efficient utilization of resources and accelerated training processes in network traffic classification tasks.

However, deep learning for network traffic classification in IoT also presents some challenges and limitations. Deep learning models often require substantial computational resources and training time, which can be a concern in resource-constrained IoT devices with limited processing power and energy. Additionally, the need for extensive labeled datasets for training deep learning models may be challenging to obtain in some IoT contexts. Furthermore, deep learning models are often regarded as black boxes, making it difficult to interpret and explain their decision-making processes, which can be problematic in scenarios where explainability and accountability are crucial. The implications of deep learning in network traffic classification for IoT are significant. Accurate classification of network traffic can enable efficient resource allocation, traffic management, and quality of service (QoS) mechanisms in IoT systems. It facilitates the identification and prioritization of critical applications and services, leading to improved user experience and overall system performance. Additionally, deep learning-based traffic classification plays a vital role in IoT security, enabling the detection and mitigation of network anomalies, intrusion attempts, and malicious activities. It empowers network administrators and security professionals to enforce appropriate security policies and protect IoT infrastructure from potential threats and vulnerabilities.

Deep learning offers a promising approach for network traffic classification in the IoT domain. Its ability to automatically learn complex patterns, handle large-scale datasets, and adapt to dynamic IoT environments makes it a valuable tool for managing and securing IoT systems. However, challenges related to computational requirements, data availability, and interpretability need to be addressed. As research progresses, deep learning techniques in network traffic classification are expected to evolve, leading to further advancements in IoT network management, security, and resource optimization. Future research efforts should focus on addressing the limitations of deep learning models in IoT contexts and exploring techniques to optimize their performance for real-time and resource-constrained IoT applications.

## 3 RELATED WORK

In several surveys published in recent years researchers have addressed various challenges and proposed solutions in the field of network traffic analysis. One survey [8] provides a comprehensive review of network traffic classification techniques, discussing their implementations, advantages, and limitations. The authors emphasize the importance of network traffic classification for purposes like quality of service, lawful interception, preventing choke points, and identifying malicious behavior. They suggest integrating multilayer classification models to improve accuracy and overcome limitations, while also proposing the exploration of supervised learning, evaluation of models in real environments with high-scale traffic, multi-source traffic classification, and combining statistical feature analysis with deep learning.

Another survey [38] focuses on the challenges posed by the increasing adoption of network traffic encryption. It reviews existing literature on solutions for analyzing encrypted network traffic and highlights the constraints faced by traditional deep packet inspection systems. Ongoing efforts within the research community to overcome these limitations are acknowledged, and future research directions in encrypted traffic analysis and processing are identified. Additionally, a survey [22] highlights the use of image processing techniques in network traffic analysis, particularly in the context of efficient and automated analysis. The authors discuss the application of artificial intelligence and deep neural networks in analyzing network traffic image data and note the existing challenges in determining suitable image representations for specific environments and conditions.

Furthermore, a paper [16] addresses the lack of standardization in network traffic analysis, proposing the use of pcapML, an open-source system, for standardized dataset curation and usage. The authors also introduce pcapML benchmarks to track the progress of network traffic analysis methods and demonstrate the impact of standardization on reproducibility and innovation in the field. Collectively, these publications provide valuable insights into network traffic analysis, covering topics such as classification techniques, the impact of encryption, image processing applications, and the importance of standardization.

Several publications have explored different aspects of Internet of Things (IoT) security and classification techniques. One survey [42] focuses on identifying and profiling IoT devices, discussing various profiling methods and their categorization based on security perspectives. The survey emphasizes the need for specialized tools to monitor IoT devices and highlights the importance of accurate device identification for security.

Another article [45] delves into the use of machine and deep learning techniques for IoT security intelligence. It provides an overview of IoT security challenges, traditional security limitations, and the extraction of insights from raw data to protect IoT devices against cyber-attacks. The article highlights the significance of selecting appropriate models for IoT security and identifies future research directions.

In the context of IoT platform support, a survey [43] explores FPGA-based implementations of classification techniques to process large datasets efficiently. It discusses different classification techniques, existing FPGA implementations, challenges, and optimization strategies. The relevance to IoT lies in the need for high-speed classification in IoT applications, and FPGA-based implementations can enhance performance.

Additionally, a review paper [51] presents a comprehensive survey on machine learning and deep learning perspectives of Intrusion Detection Systems (IDS) for IoT. It discusses IDS placement, analysis strategies, intrusion categories, and the application of machine learning and deep learning for attack detection. The paper emphasizes the importance of robust IDS solutions for IoT security and suggests future research directions. Together, these publications contribute valuable insights into IoT device profiling, security intelligence, FPGA-based implementations, and IDS perspectives, addressing security challenges and identifying avenues for further exploration.

Publications have also explored the applications of deep learning techniques in network traffic classification, monitoring, and intrusion detection in the context of community networks and modern communication systems. One review [32] focuses on traffic classification in community networks and highlights the effectiveness of deep learning in addressing the challenges posed by encrypted traffic. The paper discusses various approaches and techniques used in traffic classification and emphasizes the need for near-real-time fine-grained classification in community networks. It suggests exploring packet-based or flow-based methods and utilizing deep learning models such as CNNs and LSTMs, while considering computational resource usage and training time.

Another survey [1] provides a comprehensive overview of deep learning techniques in Network Traffic Monitoring and Analysis (NTMA) for systems like IoT and cellular networks. It covers a wide range of applications including traffic classification, prediction, fault management, and network security. The paper discusses the advantages, disadvantages, challenges, and future research directions in the field of deep learning for NTMA. In the context of IoT intrusion detection, a research study [21] explores the implementation of top artificial intelligence deep learning techniques using the IoT-23 dataset. The study develops various neural network models and identifies CNN, GANs, and multilayer perceptron as achieving the highest accuracy scores with minimal loss function and execution time. The research highlights the effectiveness of deep learning models, particularly CNN, in detecting IoT anomalies and reducing attacks. Collectively, these publications contribute valuable insights into the applications of deep learning for network traffic classification, monitoring, and intrusion detection in various contexts, offering guidance for future research in the field.

## 4  METHODOLOGY

To conduct this survey, we performed a literature search using relevant search queries. The search queries included terms such as "Internet of Things," "IoT," "network traffic classification," "deep learning," and "deep neural networks." We utilized multiple search queries to ensure a thorough exploration of the literature. The queries used were ["Internet of Things," "IoT," "network traffic classification," "deep learning," "deep neural networks"], ["IoT," "network traffic classification," "deep learning," "deep neural networks," "Internet of Things"], and ["Internet of Things," "IoT," "network traffic classification," "deep learning," "deep neural networks"].

We searched various academic databases, including but not limited to IEEE Xplore, ACM Digital Library, and Google Scholar. The search results were filtered based on relevance to the topic and inclusion criteria. After obtaining the relevant papers, we reviewed them to extract information on the deep learning approaches employed for network traffic classification in the context of IoT. We focused on techniques, algorithms, architectures, and datasets utilized in the studies. By analyzing the collected information, we categorized and summarized the different deep learning approaches used for network traffic classification in the IoT. We identified common trends, challenges, and advancements in the field. The methodology employed in this survey aimed to provide a comprehensive overview of the current state of research on deep learning approaches for network traffic classification in the IoT.

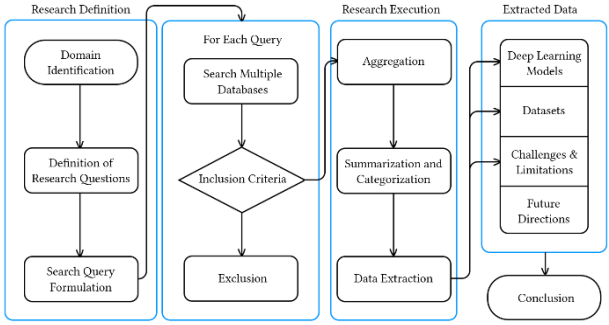

**Figure 1 Methodology for this Survey**

## 5 DISCUSSION

The literature on deep learning approaches for network traffic classification in the Internet of Things (IoT) faces several challenges. These include limited benchmarking and comparative analysis, inconsistent experimental setups, insufficient analysis of real-world challenges, inadequate focus on interpretability and explainability, and imbalanced representation of research contributions. These issues hinder the assessment, comparison, and practical implementation of deep learning models in IoT network traffic classification. Addressing these concerns would improve the reliability, applicability, and overall progress in the field.

Despite claims of being tailored for the Internet of Things (IoT), certain publications in the literature often fall short in fulfilling this assertion. While the authors purport to address IoT-specific challenges, their work frequently lacks a comprehensive understanding of the distinctive characteristics, requirements, and limitations of IoT networks. In order to bridge this gap, future research must diligently adhere to the unique demands and complexities of the IoT landscape, ensuring that proposed solutions can be effectively deployed in real-world IoT environments.

However, despite the challenges present in the literature, we have successfully tabulated and consolidated the existing research in this field. Through meticulous analysis, we have derived noteworthy insights and findings.

The table below presents a summary of the literature, organizing the publications based on their reference, year of publication, proposed solution or approach, dataset(s) utilized in their studies, and the specific purpose of network traffic classification. This tabulated information provides a concise overview of the publications, allowing researchers and practitioners to gain insights into the proposed solutions, datasets employed, and the specific objectives of network traffic classification addressed in each study.

**Table 1: A Selection of Recent Deep Learning Models for IoT Network Traffic Classification**

| Ref | Year | Proposed Solution | Dataset(s) | Classification Purpose |
|---|---|---|---|---|
| [17] | 2020 | Deep CNN | Moore dataset | Application based Classification |
| [2] | 2020 | Deep CNN | NSL-KDD | Binary & Multiclass Malicious Traffic Classification |
| [20] | 2020 | D-PACK: CNN + Autoencoder for auto-profiling | USTC-TFC2016, Mirai-RGU, Mirai-CCU | Malicious Traffic Classification |
| [10] | 2020 | ResNet-50 | Custom (converted to binvis) | Malicious Traffic Classification |
| [57] | 2020 | Mini-Batch Gradient Descent with an Adaptive Learning Rate and Momentum | Custom | Anomaly Detection |
| [47] | 2021 | Cost sensitive CNN | ISCX VPN-nonVPN | Application based Classification |
| [54] | 2021 | ByteSGAN: A Semi-Supervised Generative Adversarial Network | ISCX VPN-nonVPN + Custom | Application based Classification |
| [55] | 2021 | Logarithmic Neural Network (LOGNN) | NSL-KDD and UNSW-NB15. | Intrusion Detection |
| [15] | 2021 | Based on Deep Transfer Learning | USTC-TFC2016 | 5G Network Traffic Classification |
| [5] | 2021 | Deep Autoencoder | Bot-IoT | Botnet Detection |
| [39] | 2021 | Stacked RNN (SRNN) | Bot-IoT | Botnet Detection |
| [27] | 2021 | TSCRNN | ISCXTor2016 | Encrypted Traffic Classification |

| Ref | Year | Proposed Solution | Dataset(s) | Classification Purpose |
|---|---|---|---|---|
| [14] | 2021 | Multiclass Feed Forward Neural Network mFNN | Bot-IoT | Intrusion Detection |
| [25] | 2022 | Multi-task lEarning Model with hyBrid dEep featuRes (MEMBER): CNN with attention | Bot-IoT, UNSW-NB15, CICIDS2017, ISCX2012 | Intrusion Detection |
| [48] | 2022 | Stack Ensembled Meta-Learning-Based Optimized Classification Framework | UNSW-NB15 | Multiclass Traffic Classification |
| [58] | 2022 | Deep Neural Network | Custom + Ton-IoT | Intrusion Detection |
| [26] | 2022 | LSTM Autoencoder in an unsupervised manner | NEU-SNS-intl-IoT | Zero-Day Attack Detection |
| [53] | 2022 | Multimodal Autoencoder + BiLSTM | Custom IoMT-blockchain dataset | Anomaly Detection |
| [28] | 2022 | Packet Graph-Vector Transformer + CNN + Multi-Dimensional CNN | UNSW-NB15 + Custom | Encrypted Traffic Identification |
| [52] | 2022 | LSTM, 1D-CNN, Deep Forest, RF | Custom | Multiclass Device Identification. |
| [35] | 2022 | Deep CNN | IoT-23 dataset | Malicious Traffic Classification |
| [9] | 2022 | CNN With Deep Feature Extraction | UNSW-NB15, CICIDS2017, and KDDCup99 | Real-Time Intrusion Detection |
| [19] | 2022 | 3 Models: VGG-DALNet Model, Res-DALNet Model, Alex-DALNet Model | USTC-TFC2016 | Malicious Traffic Classification |
| [37] | 2022 | Recurrent Kernel CNN-Modified Monarch Butterfly Optimization | N-BaIoT and CICIDS-2017 | Intrusion Detection |
| [30] | 2022 | 2 Stage Distillation Aware Compressed Deep CNN Models | ISCX VPN-nonVPN | Application based Classification |
| [34] | 2022 | Knowledge-Transfer-ConvLaddernet and KT-Domain-Adaptive-ConvLaddernet | USTC-TFC2016 + Custom | Malicious Traffic Classification |
| [60] | 2022 | LSTM DNN | UNSW-NB15 and Bot-IoT | Intrusion Detection |
| [56] | 2022 | Simplified Time Convolutional Network (S-TCN) | CSE-CICIDS2018 + Custom | Multi-Class Malicious Traffic Classification |
| [44] | 2022 | Confidence Measure-Based Ensemble Deep Learning Model | ISCX VPN-nonVPN | Realtime Multiclass Traffic Classification |
| [41] | 2022 | Transfer Learning with CNNs | BoT-IoT and UNSW-NB15 | Zero-Day Attack Detection |
| [18] | 2023 | Deep Subdomain Adaptation Network with Attention Mechanism (DSAN-AT) | MCFP | Malicious Traffic Classification |
| [61] | 2023 | Cost Matrix Time-Space Neural Network (CMTSNN) | ToN-IoT, BoT-IoT and ISCX VPN-nonVPN | Anomaly Detection |
| [13] | 2023 | Deep learning Multiclass classification model (EIDM) | CICIDS2017 | Anomaly-Based Intrusion Detection |
| [31] | 2023 | Deep Learning-Based Convolutional Neural Network | CICDDoS2019 | Binary And Multi-Class Classification |
| [4] | 2023 | Federated Blending Model | Edge-IIoT, InSDN | Intrusion Detection |
| [50] | 2023 | RNN + BiLSTM | BoT-IoT | Intrusion Detection |
| [11] | 2023 | Global-Local Attention Data Selection (GLADS) | ISCX VPN-nonVPN, ISCX-Tor, USTC-TFC, ToN-IoT | Multimodal Multitask Encrypted Traffic Classification |

## 5.1 Application Awareness in IoT

Application awareness in network traffic classification in IoT refers to the ability to identify and understand the specific applications or services generating the network traffic. It involves classifying traffic based on the applications or protocols being used, such as video streaming, voice over IP (VoIP), web browsing, or file transfer. Application awareness is important because it allows for more precise and granular monitoring, analysis, and control of network traffic. By accurately identifying different applications, network administrators can optimize network resources, prioritize critical applications, detect and mitigate potential security threats, and ensure a better quality of service for IoT devices and applications.

Several models have been introduced to enhance application awareness in network traffic classification for IoT. One such model proposed in [17] is a CNN-based mechanism integrated with Software-Defined Networking (SDN). The mechanism consists of three modules: traffic collection, data pre-processing, and application-awareness. Another approach presented in [47] is the Cost-Sensitive Convolutional Neural Network (CSCNN), which addresses the class imbalance problem in low-frequency traffic data identification. The CSCNN adapts misclassification costs during training, leading to improved traffic classification, particularly for low-frequency traffic data. Furthermore, a GAN-based Semi-Supervised Learning Encrypted Traffic Classification method called ByteSGAN is proposed in [54]. This approach utilizes a small number of labeled traffic samples and a large number of unlabeled ones to achieve fine-grained traffic classification. ByteSGAN outperforms other supervised learning methods like CNN in terms of traffic classification.

Compressed models have also been explored for traffic classification in IoT networks. The paper discussed in [30] introduces a two-step distillation scheme using a Network in Network (NIN) model. Channel pruning is applied to select important filters, resulting in compressed models. Knowledge Distillation (KD) is utilized to transfer soft target, relationship, and feature map information, achieving higher accuracy and significantly reducing computation resources compared to state-of-the-art CNN models.

The proposed models offer notable strengths in network traffic classification. The CNN-based mechanism with SDN integration [17] shows superior performance in terms of key metrics, indicating its effectiveness in application-awareness. The CSCNN approach [47] addresses the class imbalance problem, making it suitable for low-frequency traffic data identification. ByteSGAN [54] demonstrates superior performance in encrypted traffic classification. The compressed models [30] achieve higher accuracy with reduced computation resources, addressing the problem of unbalanced traffic types. The HetIoT-CNN IDS [31] shows high accuracy in detecting various attacks while maintaining efficiency.

However, these models also have certain limitations. The t-SNE pooling function in the CNN-based mechanism [17] introduces high computation complexity, impacting application-awareness speed. Similarly, the CCN structure model has too many hidden layers, leading to increased computation time and slower convergence speed. Future work involves improving the t-SNE function and finding an optimal CCN structure model. The ByteSGAN model [54] addresses data imbalance issues in traffic classification but faces challenges related to GAN training, such as mode collapse and instability. Prequential evaluation technology is suggested for evaluating streaming data in the context of traffic classification. While the compressed models [30] perform well, further enhancements are needed to handle the problem of unbalanced traffic types. Future studies aim to develop efficient models addressing this issue.

## 5.2 Classification of Encrypted traffic & Others

Encrypted and other unique IoT traffic classification in IoT refers to the identification and analysis of network traffic that is encrypted or uses unique protocols specific to IoT devices. It involves detecting and classifying traffic that cannot be easily inspected or understood due to encryption or unconventional protocols. This type of classification is important because it enables the monitoring and detection of potential security threats or malicious activities that may be hidden within encrypted or unique IoT traffic. By accurately classifying and analyzing such traffic, network administrators can enhance the security and integrity of IoT networks, detect anomalies, and protect against potential attacks or unauthorized access.

Researchers have proposed diverse models to address the unique challenges of network traffic classification in IoT systems. One approach utilizes deep transfer learning to overcome limited datasets and computing capabilities in 5G IoT systems [15]. This method employs weight initialization and neural network fine-tuning to enable transfer learning between different domains. Experimental results demonstrate the high accuracy achieved by models such as LeNet-5, BiT, and EfficientNet-B0, even with limited labeled data. For encrypted traffic classification in Industrial IoT (IIoT) systems, a novel approach called TSCRNN has been introduced [27]. TSCRNN considers both temporal and spatial characteristics, outperforming other classification methods based on machine learning and deep learning. This approach shows promise for various IIoT applications, including communication, network, information processing, and security. In the domain of IoT device identification, an autonomous model update framework based on packet payloads has been proposed [28]. This framework accurately identifies known device traffic while mitigating the impact of unknown interference in an open-world scenario. The 2-D CNN classifier in the framework achieves accurate classification of known IoT devices and exhibits resistance against unknown interference. GLADS is a lightweight, multitask, and deep learning-based encrypted traffic classification model designed for resource-constrained IoT environments [11]. This model incorporates the "indicator" mechanism for simultaneous feature extraction from multiple modalities. Experimental results demonstrate that GLADS outperforms existing baselines and provides insights into the relationship between input lengths and model performance. To classify time-series signals generated by IoT device network flows, a deep learning-based approach using different architectures has been proposed [52]. The DF model, based on the Deep Forest algorithm, shows competitive performance compared to 1D-CNN and LSTM models, outperforming the Random Forrest model. Future work involves exploring scenarios with concept drifts caused by events like firmware upgrades.

The proposed models offer several strengths in network traffic classification for IoT. The deep transfer learning approach [15] enables accurate classification even with limited labeled data, leveraging techniques proven effective in image classification domains. TSCRNN [27] demonstrates superior performance by considering both temporal and spatial characteristics, enhancing encrypted traffic classification in IIoT systems. The autonomous model update framework [28] accurately identifies known IoT device traffic while being resilient to unknown interference, improving network management. GLADS [11] provides a lightweight and multitask model for encrypted traffic classification in resource-constrained IoT environments. The DF model [52] achieves competitive performance without requiring feature engineering or hyper-parameter tuning.

However, these models also have limitations that should be considered. The deep transfer learning approach [15] relies on the availability of labeled data, which may still be limited in certain IoT scenarios. TSCRNN [27] would benefit from addressing the issue of unbalanced traffic datasets and exploring unsupervised and semi-supervised learning methods. The autonomous model update framework [28] should be further validated and optimized for diverse real-world scenarios. GLADS [11] needs further research to handle zero-day applications and address imbalanced traffic. The DF model [52] exhibits higher inference time compared to other models, which may impact real-time classification requirements.

### 5.3 Malicious Traffic Identification

Malicious traffic identification through network traffic classification in IoT refers to the process of detecting and classifying network traffic that exhibits malicious behavior or intent. It involves identifying patterns, signatures, or anomalies in the network traffic that indicate potential attacks, malware, or unauthorized activities. This type of identification is crucial because it enables the early detection and mitigation of security threats in IoT environments. By accurately identifying and classifying malicious traffic, network administrators can take proactive measures to protect IoT devices, networks, and data, ensuring the integrity, privacy, and security of the entire IoT ecosystem.

Researchers have proposed several models to detect and classify malicious traffic in IoT networks. IoT-IDCS-CNN [2] presents an efficient deep-learning-based system that utilizes a CNN design to achieve high detection accuracy for distinguishing normal and anomaly traffic. The system utilizes high-performance computing with Nvidia GPUs and Intel CPUs for parallel processing. The proposed system consists of three subsystems: feature engineering, feature learning, and detection/classification. Using a CNN-based design, the system achieves high detection accuracy (99.3%) for distinguishing normal and anomaly traffic and high classification accuracy (98.2%) for categorizing IoT traffic into five classes.

Another framework called D-PACK [20] incorporates traffic sampling, traffic auto-profiling using a CNN, and an unsupervised deep learning model (autoencoder) to enable early detection of malicious traffic. This system demonstrates near-perfect detection accuracy and offers the advantage of speeding up the detection process. D-PACK aims to examine a minimal number of packets and bytes to reduce processing volume. D-PACK's efficiency is highlighted by its ability to detect with only two packets per flow and 80 bytes per packet. The framework is expected to consume less flow pre-processing and detection time compared to prior works.

Additionally, deep learning models like ResNet50 [10] and VGG-DALNet [19] have been employed for identifying malicious network traffic, achieving high accuracy rates even with limited labeled samples. The paper evaluates the accuracy of three models (VGG-DALNet, Res-DALNet, and Alex-DALNet) using SSL-based MTC methods for classifying malicious traffic with limited labeled samples. VGG-DALNet performs the best among the three models, followed by Res-DALNet and Alex-DALNet.

[34] propose Multi-Task Classification (MTC) methods for secure IIoT applications, where ConvLaddernet-based MTC performs well with few labeled samples, and KT-ConvLaddernet-based MTC excels with even fewer labeled samples. DSAN-AT [18] introduces a deep transfer learning model that accurately identifies malware variant traffic at an IoT edge gateway, utilizing channel and spatial attention mechanisms for enhanced performance. Lastly, Multi-class S-TCN [56] presents an improved Temporal Convolutional Network (TCN) solution based on Deep Packet Inspection (DPI), offering high accuracy and fast detection speed for malicious traffic in IoT environments. The [35] DEMD-IoT model is proposed for IoT malware detection using deep ensemble learning architectures. One-dimensional CNNs are used instead of two-dimensional CNNs, reducing preprocessing time and computational complexity. Techniques such as Batch Normalization, Dropout, and early stopping are employed to prevent overfitting and improve outcomes. Three 1D-CNN classifiers with different settings are built to learn various patterns of IoT network traffic. The models' hyperparameters are optimized using the GridSearchCV algorithm.

The proposed models for malicious traffic detection in IoT network traffic classification offer several strengths. IoT-IDCS-CNN [2] achieves high detection and classification accuracy, surpassing existing IDS systems in the same domain. D-PACK [20] stands out for its ability to speed up detection, allowing for timely identification of malicious traffic. Deep learning models such as ResNet50 [10] and

VGG-DALNet [19] demonstrate high accuracy rates, even with limited labeled samples, making them effective in real-world scenarios. The MTC methods proposed in ConvLaddernet and KT-ConvLaddernet [34] cater to different labeled sample scenarios, showcasing their versatility. DSAN-AT [18] exhibits high performance in accurately identifying specific malware variant traffic, even with a small training dataset. Multi-class S-TCN [56] offers high accuracy, fast detection speed, and support for parallel detection, making it suitable for IoT environments. The DEMD-IoT model [35] achieves superior performance compared to state-of-the-art models through the use of one-dimensional CNNs and optimization techniques.

Despite their strengths, the proposed models also have certain limitations. IoT-IDCS-CNN [2] suggests the need for additional data collection and customization with other cyberattack datasets to enhance its performance. D-PACK [20] could benefit from further optimization methods to reduce detection delays. The deep learning models like ResNet50 [10] and VGG-DALNet [19] require labeled samples for training, which may be challenging to obtain in certain scenarios. ConvLaddernet and KT-ConvLaddernet [34] rely on pretraining models, posing practical challenges. DSAN-AT [18] requires refinement to consider adaptability, federated transfer learning, and online transfer learning for identifying specific malware variant traffic. The DEMD-IoT model [35] may experience increased execution time due to a large number of hyperparameters, and exploring parallel processing is recommended to reduce computational costs.

### 5.4 Anomaly Detection

Anomaly detection in network traffic classification in IoT refers to the process of identifying abnormal or suspicious patterns in network traffic behavior. It involves analyzing network traffic data to detect deviations from normal behavior that may indicate potential security breaches, intrusions, or abnormal activities. This type of detection is important because it helps identify and mitigate unknown or emerging threats in IoT networks. By accurately detecting anomalies in network traffic, administrators can take prompt action to investigate and respond to potential security incidents, ensuring the overall security and integrity of IoT devices, data, and infrastructure.

Researchers have proposed various models to detect anomalies in network traffic in the IoT environment. A feature extraction method using multi-model autoencoders is proposed in [53] to effectively extract and fuse features from different traffic feature subspaces. The method also introduces a multi-feature sequence anomaly detection algorithm using residual learning, showcasing good performance in anomaly detection for IoT-Blockchain traffic. In [61], the CMTSNN model is presented for multi-classification identification of encrypted abnormal traffic in IoT. This model combines BiLSTM-1DCNN for temporal and spatial feature extraction, and employs a cost penalty matrix and an improved cross-entropy loss function to address unbalanced traffic data. Experimental results demonstrate superior performance, with lower false alarm rates and higher accuracy. In [57], the HCAMBGDALRM algorithm is introduced for mining untrustworthy data from massive traffic data in Industrial IoT (IIoT). This algorithm outperforms others in terms of performance and precision, enhancing trustworthiness in networking big data in IIoT. Additionally, [13] presents the EIDM model, which utilizes deep learning to detect and classify suspicious behaviors in network flow. The model achieves high accuracy in classifying various traffic behaviors, outperforming other models in terms of accuracy and time cost.

The proposed models for anomaly detection in network traffic classification in IoT offer several strengths. The feature extraction method in [53] effectively extracts and fuses features from different traffic feature subspaces, enhancing anomaly detection performance. The CMTSNN model in [61] combines temporal and spatial feature extraction methods, leading to improved robustness and identification rates for encrypted abnormal traffic. The HCAMBGDALRM algorithm in [57] outperforms other algorithms, ensuring trustworthiness in networking big data in IIoT and supporting safety improvement. The EIDM model in [13] achieves high accuracy in classifying suspicious behaviors, surpassing other models and demonstrating efficient time cost.

Despite their strengths, the proposed models also have certain limitations. The feature extraction method in [53] requires further research to develop more flexible algorithms and study real-time anomaly detection in simulated network environments. The CMTSNN model in [61] requires exploration of adaptability to real-time flow changes and optimization to maintain identification rates while reducing model complexity. The HCAMBGDALRM algorithm in [57] provides parallel processing support but may require additional investigation to address specific IIoT scenarios. The EIDM model in [13] would benefit from exploring the distribution of learning processes across different machines for high-performance computing and enhancing the overall performance and security of the intrusion detection system.

### 5.5 Botnet Detection

Botnet detection in network traffic classification in IoT refers to the process of identifying and distinguishing network traffic associated with botnets, which are networks of compromised devices controlled by a malicious actor. It involves analyzing traffic patterns, communication behavior, and other characteristics to identify the presence of botnet-related activities. Botnet detection is crucial because it helps mitigate the risks posed by botnet attacks in IoT networks. By accurately detecting and classifying botnet traffic, network administrators can take necessary actions to isolate and mitigate the compromised devices, prevent further spread of malware, and safeguard the integrity, availability, and security of IoT networks and devices.

Researchers have proposed various models to detect and classify botnet attacks in network traffic in IoT. In [5], a deep autoencoder-based anomaly detection solution is proposed. This method shows high accuracy, precision, and recall values, indicating its effectiveness and robustness in detecting botnet attacks in IoT networks. In [39], the SRNN (Stacked Recurrent Neural Network) model is introduced for botnet detection in highly imbalanced network traffic data in a Smart Home Network (SHN) environment. The SRNN model utilizes multiple layers of RNN to learn hierarchical representations of the imbalanced network traffic data, leading to better representation learning and improved generalization ability.

The proposed models for botnet detection in network traffic classification in IoT offer several strengths. The deep autoencoder-based anomaly detection solution in [5] demonstrates high accuracy and robustness in detecting botnet attacks. It can effectively identify new or unknown threats, which is crucial in IoT environments where traditional signature-based systems may fall short. The SRNN model in [39] outperforms traditional RNN models, exhibiting better representation learning and robustness against overfitting. It excels in detecting network traffic samples from minority classes with high imbalance ratios, making it suitable for securing Smart Home Networks against complex botnet attacks.

Despite their strengths, the proposed models also have certain limitations. The deep autoencoder-based anomaly detection solution in [5] may be vulnerable to adversarial machine learning techniques, which could potentially evade the detection system. Additionally, it may not detect internal attacks that occur beyond the gateway level in IoT networks, limiting its scope. The SRNN model in [39] has longer training and response times compared to other machine learning and deep learning models, although this trade-off is considered insignificant given the large number of network traffic samples.

## 5.6 Intrusion Detection

Intrusion detection by network traffic classification in IoT involves utilizing deep learning techniques to identify and detect potential security breaches or unauthorized activities within IoT networks. Deep learning models are trained on large amounts of network traffic data to learn patterns and behaviors associated with normal and malicious network activity. This approach is crucial for IoT security as it enables real-time monitoring and proactive identification of anomalies, helping to protect IoT devices, networks, and sensitive data from cyber threats.

Several models have been proposed to detect and classify intrusions in IoT networks based on network traffic. The MEMBER framework proposed in [25] leverages multi-task learning for intrusion detection in imbalanced network scenarios. By combining statistical and packet content features, the model captures rich representations and exhibits improved generalization ability. The inclusion of a memory module and attention mechanisms further enhances its performance. In the HetIoT (Heterogeneous Internet of Things) environment, the HetIoT-CNN IDS, a deep learning-based CNN, is proposed in [31]. This IDS demonstrates high accuracy in detecting benign and DDoS attacks. It also showcases efficiency in terms of time, lightweight design, and low complexity compared to state-of-the-art IDSs. Additionally, the identification and feature extraction tool presented in [58] demonstrates effectiveness in filtering and identifying various types of network traffic.

Other advancements in intrusion detection for network traffic classification in IoT include the DL-based IDS framework proposed in [50], which utilizes a fog-cloud architecture to address computation and latency challenges. This framework demonstrates efficacy, reduced network latency, and improved detection capability. Furthermore, the federated blending model-driven IDS framework (F-BIDS) presented in [4] offers improved classification performance and reduced privacy risk through federated learning, addressing privacy concerns associated with centralized learning. The introduction of logarithmic neurons and the logarithmic neural network (LOGNN) in [55] shows promising results in intrusion detection, outperforming traditional deep learning and machine learning algorithms.

The customized feed-forward neural network introduced in [14] incorporates concepts like network embedding and transfer learning to enhance its performance. The RKCNN-MMBO model, comprising a kernel classifier and a DL-classifier, is employed for classification in the IoT intrusion detection mechanism (IDM) showcased in [37]. Additionally, the PB-DID (Packet-Based Intrusion Detection) approach presented in [60] addresses imbalance and overfitting issues in public datasets by combining standard flow, TCP, and other features. The Deep Feature Extraction (DFE) method described in  focuses on extracting more information from input data using 2D convolutions and permutations.

The proposed models for intrusion detection in network traffic classification in IoT offer several strengths. The MEMBER framework in [25] captures comprehensive and robust feature representations, leading to improved generalization ability. The HetIoT-CNN IDS in [31] is lightweight, efficient in terms of time, and less complex, making it suitable for resource-constrained IoT environments. The IoT intrusion detection system in [58] shows effectiveness in detecting intrusions and can be further enhanced by exploring different deep learning models with varied architectures. The customized feed-forward neural network in [14] leverages network embedding and transfer learning techniques, which enhance its ability to capture relevant features and improve classification accuracy. The RKCNN-MMBO model in [37] combines kernel and DL-classifiers, enabling effective classification after preprocessing. The PB-DID approach in [60] addresses imbalance and overfitting issues in public datasets, reducing the number of features required for identifying malicious traffic. The DFE method in [9] enhances classification accuracy by extracting more information from input data while minimizing the computational requirements, making it suitable for real-time intrusion detection in IoT devices with limited processing capabilities.

While the proposed models demonstrate strengths, they also have certain limitations. The MEMBER framework in [25] requires further research to detect multi-stage attacks with long time spans and address the impact of highly stealthy stealing attacks and adversarial attacks on model performance and robustness. The HetIoT-CNN IDS in [31] focuses on a specific environment and should explore additional models, such as Recurrent Neural Networks (RNN), for detecting and predicting DDoS attacks. The IoT intrusion detection system in [58] would benefit from expanding the dataset to include other IoT protocols, such as the MQTT protocol, for comprehensive intrusion detection coverage. The customized feed-forward neural network in [14] demonstrates subpar performance in classifying specific attack subcategories, highlighting the need for further improvement. The IDM utilizing the RKCNN-MMBO model in [37] requires reliability testing against severe attacks to ensure robust intrusion detection capabilities. The PB-DID approach in [60], although effective in addressing imbalance and overfitting, should explore the generalization of its methodology to diverse IoT environments and datasets. The DFE method in [9] could benefit from additional research to enhance the classification of minority classes by optimizing the permutation process.

To advance intrusion detection in IoT networks through network traffic classification, several areas warrant further exploration. The proposed models can be enhanced by incorporating long-short term memory networks to leverage timestamp information and header fields for differentiating attack subcategories [14]. Reliability measurement against severe attacks and accuracy evaluation of DL techniques can be conducted to strengthen the IDM utilizing the RKCNN-MMBO model [37]. Future work should also focus on extending the applicability of the PB-DID approach to a wider range of IoT environments and expanding the dataset coverage to include additional IoT protocols [60]. For the DFE method [9], future research can concentrate on optimizing the permutation process as an optimization problem, enabling improved classification of minority classes.

### 5.7 Zero-Day Attack Detection

Zero-day attack detection by network traffic classification in IoT involves identifying and mitigating previously unknown or undisclosed vulnerabilities and attack techniques that exploit these vulnerabilities. Unlike traditional intrusion detection systems that rely on known attack patterns, zero-day attack detection employs advanced machine learning algorithms to analyze network traffic data and identify anomalous behavior that may indicate a new or unknown attack. This approach is crucial for IoT security as it provides an additional layer of defense against emerging threats, ensuring the timely detection and prevention of attacks that could exploit vulnerabilities that have not yet been patched or addressed by security updates. By proactively identifying and mitigating zero-day attacks, organizations can significantly reduce the potential damage caused by these advanced threats and safeguard their IoT infrastructure and sensitive data.

Multiple models have been proposed for zero-day attack detection in IoT networks through network traffic classification. The intrusion detection framework introduced in [41] utilizes transfer learning and model refinement techniques to improve detection accuracy in limited and imbalanced datasets. The ADRIoT framework presented in [26] adopts an edge-assisted architecture and incorporates a multiedge collaborative mechanism to enable prompt detection of IoT-based attacks.

The proposed models offer several strengths in the context of zero-day attack detection in network traffic classification in IoT. The intrusion detection framework in [41] demonstrates excellent accuracy and low false positive rates, even for novel zero-day attack families. The utilization of transfer learning and network fine-tuning improves detection rates and outperforms previous deep learning-based intrusion detection systems. The ADRIoT framework in [26] leverages an edge-assisted architecture, enabling the anomaly detection module to run closer to the data source for real-time detection. The incorporation of a multiedge collaborative mechanism enhances the resource utilization on the edge, supporting efficient and effective detection of a wide range of IoT-based attacks.

Despite their strengths, the proposed models also face certain limitations. The intrusion detection framework in [41] primarily focuses on IoT network traffic from the UNSW-NB15 dataset, necessitating further evaluation on real data from diverse IoT networks. Future research should extend the framework to detect other types of zero-day attacks and assess its performance on lightweight IoT devices with real IoT network traffic. The ADRIoT framework in [26] relies on unsupervised anomaly detection, which may limit its ability to detect novel zero-day attacks with high accuracy. Additional research is needed to enhance the framework's capability to handle emerging and sophisticated zero-day attack patterns.

To advance zero-day attack detection in IoT networks through network traffic classification, several areas warrant further exploration. The intrusion detection framework proposed in [41] can be extended to incorporate real data from IoT networks, allowing for a more comprehensive evaluation of its effectiveness and robustness. Future work should also focus on enhancing the framework's ability to detect diverse zero-day attack types, addressing the challenges posed by lightweight IoT devices, and exploring techniques for handling real-time IoT network traffic. In the case of the ADRIoT framework [26], future research should consider incorporating supervised learning methods to improve the accuracy of zero-day attack detection and further refine the collaborative mechanism to optimize resource utilization on the edge.

### 6  CONCLUSIONS AND FUTURE DIRECTIONS

Advancements in network traffic classification for IoT using deep learning techniques have shown promising results in various areas, including application awareness, accuracy improvement, malicious traffic detection, anomaly detection, botnet detection, intrusion detection, and zero-day attack detection. These advancements have brought several strengths to the field. For instance, models such as CSCNN, ByteSGAN, deep transfer learning, TSCRNN, GLADS, DF model, and others demonstrate superior performance, versatility, and fast detection speed. They exhibit strengths like effective feature extraction, robustness in handling encrypted traffic, enhanced trustworthiness, and improved representation learning.

However, these models also face certain limitations that need to be addressed. Computation complexity, convergence speed, GAN training challenges, unbalanced traffic types, limited labeled data, unbalanced traffic datasets, real-world validation, handling zero-day applications, inference time, optimization requirements, and computational complexities are some of the weaknesses associated with these models. Furthermore, vulnerabilities to adversarial machine learning techniques, limited detection scope, longer training and response times, and challenges in handling multi-stage attacks and adversarial attacks are additional limitations.

To overcome these limitations and further enhance network traffic classification in IoT environments, future research efforts should focus on exploring additional techniques. Recurrent neural networks, reinforcement learning, transfer learning, federated learning, parallel processing, real-time anomaly detection, high-performance computing, and supervised learning for labeling anomalies are some of the directions that should be pursued. Additionally, refinement of models, exploration of new architectures and techniques, expansion of datasets, addressing privacy and explainability concerns, differentiation of attack subcategories, reliability testing, generalization to diverse environments, optimization of classification for minority classes, and evaluation on real IoT network data are key areas for future research. By addressing these limitations and pursuing these avenues of research, advancements in network traffic classification in IoT using deep learning techniques can lead to improved security, privacy, trustworthiness, and defense against various threats and vulnerabilities.